\documentclass[reprint,prl,amsmath,amssymb,aps,showpacs,preprintnumbers]{revtex4-1}
\usepackage{graphicx} 
\usepackage{datetime}
\usepackage{bm} 
\usepackage{xcolor}

\begin{document}
\title{Local Density of States for Nanoplasmonics}
\author{Tigran V. Shahbazyan}
\affiliation{
Department of Physics, Jackson State University, Jackson, Mississippi
39217 USA}

\date{\today,\,\,\currenttime}

\begin{abstract} 
We obtain the local density of states (LDOS) for any nanoplasmonic system in the frequency range dominated by a localized surface plasmon. By including the Ohmic losses in a consistent way, we show that the plasmon LDOS is proportional to the local field intensity normalized by the  absorbed power. We obtain explicit formulas for the energy transfer (ET) between quantum emitters and plasmons as well as between donors and acceptors situated near a plasmonic structure. In the latter case, we find that the plasmon-assisted ET rate is proportional to the LDOS product  at the donor and acceptor positions, obtain, in a general form, the plasmon ET enhancement factor, and establish the transition onset between F\"{o}rster-dominated and plasmon-dominated ET regimes.  
\end{abstract}
\maketitle


%
The rapid advances in nanoplasmonics of the past decade opened up possibilities for energy concentration and transfer at length scales well below the diffraction limit \cite{stockman-review}. Optical interactions between  dye molecules or semiconductor quantum dots,  hereafter referred to as quantum emitters (QEs),  and localized plasmons in metal-dielectric composite nanostructures  underpin major phenomena in plasmon-enhanced spectroscopy,  including surface-enhanced Raman scattering  \cite{sers}, plasmon-assisted fluorescence \cite{feldmann-prl02,novotny-prl06,sandoghdar-prl06} and energy transfer  \cite{lakowicz-jf03,andrew-science04,lunz-nl11},  strong QE-plasmon coupling \cite{bellessa-prl04,sugawara-prl06,fofang-nl08}, and the plasmonic laser (spaser) \cite{bergman-prl03,noginov-nature09,zhang-nature09}. The interaction of a QE,  located at $\bm{r}$, with electromagnetic modes is characterized by  the local density of states (LDOS)  $\rho(\omega,\bm{r})=(2\omega/\pi c^{2}) \text{Im} [\text{Tr} \bar{\textbf{G}}(\omega;\bm{r},\bm{r})]$, where $\bar{\textbf{G}}(\omega;\bm{r},\bm{r}')$ is the electromagnetic Green dyadic  and $c$ and $\omega$ are speed and frequency of light, which represents  the number of modes in unit volume and frequency interval \cite{novotny-book}. In particular, the LDOS quantifies the Purcell  enhancement of spontaneous emission  by a QE  in a photonic environment \cite{purcell-pr46}, e.g.,  near metal  surfaces \cite{dereux-jcp00,greffet-prb03,polman-prb09,carminati-prl10}, metamaterials \cite{noginov-ol10,poddubny-prb12}, or plasmonic nanostructures \cite{carminati-oc06,negro-oe10,lalanne-prl13,carminati-ssr15}.  A closely  related quantity, the cross density of states (CDOS)  $\rho(\omega;\bm{r},\bm{r}')=(2\omega/\pi c^{2}) \text{Im} [\text{Tr} \bar{\textbf{G}}(\omega;\bm{r},\bm{r}')]$,  describes spatial correlations, e.g., due to indirect coupling between QEs \cite{carminati-prl13}. While for high-symmetry systems, such as flat surfaces or spherical particles, the electromagnetic LDOS is known, its evaluation for  general-shape systems  presents a rather challenging task.  A photon emission by a QE involves all system eigenmodes that define the continuum of final states \cite{hill-pra90,hughes-ol12}, so that, in open systems,   the calculations of the LDOS and CDOS  rely on carefully defined quasinormal modes \cite{lalanne-pra14,hughes-pra15}.

At the same time, nanoplasmonic systems support a host of phenomena  that are underpinned by \textit{nonradiative} plasmon-assisted transitions. For example,   energy transfer (ET) between QEs and plasmons, whose frequencies are tuned to resonance, is the key process in many plasmonics applications \cite{lakowicz-ab01,duyne-nm06}. The magnitude and range of the F\"{o}rster ET  between a donor and an acceptor  near a plasmonic structure is strongly enhanced by the plasmon-mediated ET channel \cite{nitzan-cpl84,dung-pra02,martin-pra03,pustovit-prb11}, while the role  of the LDOS in the enhancement  mechanism is a subject of ongoing debate \cite{nakamura-prb06,polman-prb05,carminati-prb11,enderlein-ijms12,blum-prl12,rabouw-nc14,wenger-nl14,noginov-fd14,vos-njp16}. Examples of coherent  plasmon-assisted processes  include strong QE-plasmon coupling  \cite{manjavacas-nl11,garcia-prl14} and  the spaser  \cite{stockman-jo10}.  These  phenomena hinge on the QEs' coupling to resonant plasmon modes that is  characterized by the \textit{plasmon} LDOS (or CDOS), which, in general, can be obtained from the electromagnetic LDOS in the near-field limit. On the other hand, in the frequency region dominated by a  localized plasmon mode, one expects  the plasmon LDOS to be  determined \textit{directly} by the mode local field. At the same time, for the system  size below the diffraction limit, the plasmon decay is mainly due to the Ohmic losses in metal, while radiation plays a relatively minor role \cite{stockman-review}. Therefore, any accurate theory the for plasmon LDOS must rely on the consistent treatment of  Ohmic losses.

Here, we develop a theory for the plasmon LDOS (and CDOS)  for any nanoplasmonic system characterized by a local dielectric function  $\varepsilon (\omega,{\bf r})=\varepsilon' (\omega,{\bf r})+i\varepsilon'' (\omega,{\bf r})$. Specifically, we show that  for $\omega$  near the plasmon  frequency $\omega_{n}$, the LDOS has a universal form
\begin{equation}
\label{ldos-main}
\rho(\omega_{n},\bm{r})=\frac{2}{\pi\omega_{n}}\frac{|\textbf{E}_{n}(\bm{r})|^{2}}{\int \! dV \varepsilon'' |\textbf{E}_{n}|^{2} },
\end{equation}
where $\textbf{E}_{n}(\bm{r})$ is the  local field determined by the Gauss law $\bm{\nabla}\cdot \left [\varepsilon' (\omega_{n},\bm{r}) \textbf{E}_{n}(\bm{r})\right ]=0$, and integration is carried over the system volume.  The plasmon LDOS is proportional to the local field intensity  normalized by the absorbed power. The derivation of Eq.~(\ref{ldos-main}), outlined below, involves a consistent treatment of the Ohmic losses, which determine the  plasmon decay rate $\gamma_{n}$, and implies a well-defined plasmon  mode with quality factor $q_{n}=\omega_{n}/\gamma_{n}\gg 1$. 
Within this approach, we  obtain general formulas for the QE-plasmon ET rates and for the donor-acceptor F\"{o}rster ET rate near  any plasmonic structure.  In the latter case, the  rate is  proportional to the  LDOS product at the donor   and acceptor   positions. We   derive the plasmon ET enhancement factor and establish a general condition that governs the transition between F\"{o}rster-dominated and plasmon-dominated ET regimes. Finally, for an ensemble of QEs coupled to a resonant plasmon mode,  we derive the cooperative ET rate in terms of the ET rates for individual QEs.

\textit{Theory}.---We consider a metal-dielectric nanostructure supporting localized plasmon modes that is characterized by  dielectric function  $\varepsilon (\omega,\bm{r})=1+4\pi\sum_{i}\chi_{i}(\omega,\bm{r})$, where $\chi_{i}(\omega,\bm{r})=\Theta_{i}(\bm{r}) [\varepsilon_{i}(\omega)-1]/4\pi$ are the local susceptibilities; $\Theta_{i}(\bm{r})$ is 1 in the region $V_{i}$ with dielectric function $\varepsilon_{i}$ and is 0 outside of it. We assume that only in \textit{metallic} regions are the dielectric functions $\varepsilon_{m}(\omega)$  dispersive and complex and that the retardation effects are unimportant. The susceptibilities $\chi_{i}$ define the  polarization vector ${\bf P}(\bm{r})=\sum_{i}\chi_{i}(\omega, {\bf r}){\bf E}(\bm{r})$, where ${\bf E}=- \bm{\nabla}\Phi$ is the local field and $\Phi(\bm{r})$ is the corresponding potential.

Our goal is to derive the plasmon Green  function and, hence, the LDOS by including, in a consistent way, the Ohmic losses that give rise to the  plasmon decay rate $\gamma_{n}$. We assume that plasmon modes are well defined, i.e., $q_{n}=\omega_{n}/\gamma_{n}\gg 1$, and adopt a perturbative approach with respect to $1/q_{n}$. We  start with  the self-consistent microscopic equation for the potential $ \Phi(\bm{r})$ \cite{mahan-book}:
\begin{equation}
\label{rpa}
\Phi(\bm{r}) = \varphi(\bm{r}) + \int dV_{1} dV_{2}u(\bm{r} -\bm{r}_{1}) P(\bm{r}_{1}, \bm{r}_{2}) \Phi(\bm{r}_{2}),
\end{equation}
where $ \hat{P}= \hat{P}^{\prime } + i \hat{P}^{\prime \prime}$ is the electron polarization operator, $u(r) =1/r$ is the Coulomb potential (we set the electron charge to unity), and $ \varphi (\bm{r})$ is an external potential.  The system eigenmodes are described by the homogeneous part of Eq.~(\ref{rpa}), which we write as $(\Delta +4\pi \hat{P})\Phi=0$, where we used that $\Delta u(\bm{r} -\bm{r}')= -4\pi\delta(\bm{r} -\bm{r}')$. The  operator $\hat{P}$ is related to the polarization vector ${\bf P}$ via the induced charge density: $\rho(\bm{r})=\int d{\bm r}'P(\bm{r}, \bm{r}') \Phi(\bm{r}')=-\bm{\nabla}\cdot {\bf P}(\bm{r})$. In the \textit{local} case, we have $\bm{\nabla} \cdot{\bf P}(\bm{r}) =\sum_{i}\bm{\nabla} \cdot [\chi_{i}(\bm{r}){\bf E}(\bm{r})]$, and the polarization operator takes the form
\begin{align}
\label{local}
P(\omega;{\bm r}, {\bm r}') =  \sum_{i}\bm{\nabla}\cdot \left[ \chi_{i}(\omega,{\bm r})  \bm{\nabla} \delta ({\bm r} - {\bm r}')\right].
\end{align}
We now introduce eigenfunctions $\Phi_{n}({\bm r})$ and  eigenvalues $\lambda_{n}(\omega)$ of the \textit{real} part of polarization operator as
\begin{equation}
\label{real}
4\pi\hat{P}'\Phi_{n}  \equiv 4\pi\sum_{i}\bm{\nabla} \cdot\left(\chi'_{i}  \bm{\nabla} \Phi_{n}  \right )= \lambda_{n} \Delta \Phi_{n}.
\end{equation}
Since $\Phi_{n}({\bm r})$ are harmonic in each region and continuous at the interfaces, they must be regular inside the nanostructure and decay sufficiently fast outside of it. Note that this approach resembles the eigenvalue problem in binary systems \cite{stockman-review,kociak-prb12}, but with the key difference that here the eigenvalues depend on the system dielectric function, allowing us to include the losses in a consistent way. From Eq.~(\ref{real}),  the  mode orthogonality follows: $\int\! dV \textbf{E}_{m}\cdot\textbf{E}_{n}=\delta_{mn} \int\! dV  \textbf{E}_{n}^{2}$. Note that the eigenfunctions of $\hat{P}'$ can always be chosen real.  Using Eq.~(\ref{real}), the eigenvalues are  found as $\lambda_{n}=4\pi \langle n|\hat{P}'|n\rangle/\langle n|\hat{\Delta}|n\rangle$. To find eigenfrequencies $\omega_{n}$, we write this expression as
\begin{equation}
\label{lambda}
1+\lambda_{n}(\omega)
=\sum_{i}\varepsilon'_{i}\,\frac{\int\! dV_{i}\textbf{E}_{n}^{2}}{\int\! dV\textbf{E}_{n}^{2}}
= \frac{\int\! dV \varepsilon' (\omega,{\bm r})\textbf{E}_{n}^{2}}{\int\! dV\textbf{E}_{n}^{2}}.
\end{equation}
For $\omega=\omega_{n}$, the right hand side of Eq.~(\ref{lambda}) vanishes due to  Gauss's law, and  so $\omega_{n}$ are found from   $\lambda_{n}(\omega_{n})=-1$.

In the presence of Ohmic losses,  the mode eigenfrequencies acquire an imaginary correction, $\omega'_{n}=\omega_n - i \gamma_{n}/2$, which can be found by  including the imaginary part of polarization operator $\hat{P}''$ in Eq. (\ref{real}). For  $q_{n}=\omega_{n}/\gamma_{n}\gg 1$, the correction $\delta\lambda_{n}$ to the eigenvalue is small and, in the first order in $1/q_{n}$, the eigenfunctions are unchanged. The new eigenfrequency  condition   reads $1+ \lambda_{n} +\delta \lambda_{n}  =0$, where $\delta \lambda_{n}=i\lambda_{n}\langle n|\hat{P}''|n\rangle/\langle n|\hat{P}'|n\rangle$. Using the expansion   $\lambda_n(\omega'_n)  = \lambda_{n}(\omega_n) - i (\gamma_{n}/2) \left[\partial\lambda_n(\omega_n)/\partial \omega_{n}\right]$ together with  $\lambda_{n}(\omega_{n})=-1$, we finally obtain  the mode decay rate as
\begin{equation}
\label{gamma}
\gamma_{n} = -2  \left (\frac{\partial \lambda_n}{\partial\omega_n}\right )^{-1} \frac{\langle n|\hat{P}''|n\rangle}{\langle n|\hat{P}'|n\rangle} = \frac{Q_{n}}{U_{n}},
\end{equation}
where we introduced the mode energy
\begin{align}
\label{energy}
U_{n}= \frac{\omega_{n}}{2}\frac{\partial \lambda_n}{\partial \omega_{n}} 
\langle n|\hat{P}'|n\rangle
=-\frac{\omega_{n}}{2}\frac{\partial \lambda_n}{\partial \omega_{n}} \,
\text{Re} \! \int  \! dV\textbf{E}_{n}\cdot\textbf{P}_{n},
\end{align}
and the absorbed power
\begin{align}
\label{power}
Q_{n}= - \omega_{n}\langle n|\hat{P}''|n\rangle
=\omega_{n}\text{Im}  \! \int  \!  dV \textbf{E}_{n}\cdot\textbf{P}_{n}.
\end{align}
Note that although the eigenstates and eigenvalues in Eq. (\ref{real}) are defined  for a local form of $\hat{P}'$, the corrections $\delta\lambda_{n}$, originating from $\hat{P}''$, may include nonlocal effects as well. In Eq. (\ref{energy}), the $\omega$ dependence of $\lambda_{n}$ comes from the metallic regions, i.e., $\partial \lambda_n/\partial \omega_{n}=\sum_{m}(\partial \lambda_n/\partial \varepsilon'_{m})(\partial \varepsilon'_{m}/\partial \omega_{n})$, and using $\textbf{P}_{n}=\textbf{E}_{n}[\varepsilon(\omega_{n},\bm{r})-1]/4\pi$, where the first term's contribution vanishes to  due to  Gauss's law, we write
\begin{equation}
\label{energy1}
U_{n}=\frac{\omega_{n}}{8\pi}\sum_{m}\frac{ \partial \varepsilon'_{m}}{\partial \omega_{n}}
\frac{\partial \lambda_n}{\partial \varepsilon'_{m}} \int\! dV\textbf{E}_{n}^{2}.
\end{equation}
Then, using  $ (\partial \lambda_n/\partial \varepsilon'_{m}) \int\! dV\textbf{E}_{n}^{2} = \int\! dV_{m}\textbf{E}_{n}^{2}$ [see Eq. (\ref{lambda})],  we recover the  usual expression for the mode energy  \cite{landau},
\begin{align}
\label{energy-LL}
U_{n}=\frac{\omega_{n}}{8\pi}\sum_{m} \frac{\partial \varepsilon'_{m}}{\partial \omega_{n}} \! \int \!dV_{m}\textbf{E}_{n}^{2}
= \!\int \! \frac{dV}{8\pi} \frac{\partial (\omega_{n}\varepsilon')}{\partial \omega_{n}}\textbf{E}_{n}^{2}.
\end{align}
Similarly, the absorbed power (\ref{power}) takes the form
\begin{equation}
\label{power-local}
Q_{n}=\frac{\omega_{n}}{4\pi}\!\int \! dV \varepsilon''(\omega_{n},\bm{r}) \textbf{E}_{n}^{2} (\bm{r})+ Q_{n}^{nl},
\end{equation}
where $Q_{n}^{nl}$ includes nonlocal contributions, e.g., due to electron-hole pairs excitation near the metal-dielectric interface \cite{kirakosyan-16}. Here we consider the local case only and disregard $Q_{n}^{nl}$ in what follows. The integrals in Eqs. (\ref{energy-LL}) and (\ref{power-local}) are, in fact, carried over the metallic regions, and, for a single metallic region, we recover the plasmon bulk decay rate: $\gamma_{n}=2\varepsilon''_{m}(\omega_{n})/[\partial \varepsilon'_{m}(\omega_{n})/\partial \omega_{n}]$.

We now turn to  Green's function $\hat{G}$ for   potentials, satisfying $(\Delta +4\pi \hat{P})G(\bm{r},\bm{r}')=-4\pi\delta (\bm{r}-\bm{r}')$, which we split into Coulomb and plasmon terms as $\hat{G}=\hat{u}+\hat{G}_{p}$,  where the latter satisfies  $(\Delta +4\pi \hat{P})\hat{G}_{p}=-4\pi \hat{P}\hat{u}$. We expand $\hat{G}_{p}$ over the eigenstates of $\hat{P}'$ as $G_{p}(\omega;\bm{r},\bm{r}')=\sum_{n}\! G_{n}^{p} (\omega)\,\Phi_{n}(\bm{r}) \Phi_{n}(\bm{r}')$, where the coefficients
\begin{equation}
\label{Gn}
G_{n}^{p}(\omega)=  \frac{\lambda_{n}(\omega)}{ \langle n|\hat{P}'|n\rangle}
\frac{ \lambda_{n}(\omega)+ \delta \lambda_{n}(\omega) }{1+ \lambda_{n}(\omega)+ \delta \lambda_{n}(\omega) },
\end{equation}
exhibit plasmon resonances. Near plasmon resonance at $\omega_{n}$, expanding $\lambda_{n}(\omega)=\lambda_{n}(\omega_{n})+(\partial \lambda_{n}/\partial \omega_{n})(\omega-\omega_{n})$ and using Eqs.~(\ref{energy})-(\ref{energy-LL}), we obtain  $G_{n}^{p}=g_{n}/(\omega-\omega_{n}+i\gamma_{n}/2)$, where $g_{n}=\omega_{n}/2U_{n}$ is the oscillator strength reflecting the fact that it is $U_{n}$, rather than $\hbar\omega_{n}$,  that is the mode energy in a dispersive medium \cite{landau}. Similarly, the Green dyadic $\bar{\textbf{D}}(\omega;\bm{r},\bm{r}')=\bm{\nabla}\otimes \bm{\nabla}'G(\omega;\bm{r},\bm{r}')$,  which matches the near-field limit of $(-4\pi\omega^{2}/c^{2})\bar{\textbf{G}}(\omega;\bm{r},\bm{r}')$,  is also a sum of  Coulomb and plasmon terms,  $\bar{\textbf{D}}=\bar{\textbf{D}}_{0}+\bar{\textbf{D}}_{p}$. For well-resolved   modes, the plasmon Green dyadic $\bar{\textbf{D}}_{p}$ is dominated by the resonant mode, and we finally obtain
\begin{equation}
\label{dyadic}
\bar{\textbf{D}}_{p}(\omega;\bm{r},\bm{r}') = \frac{\omega_{n}}{2 U_{n}}\frac{\textbf{E}_{n}(\bm{r})\otimes \textbf{E}_{n}(\bm{r}')}{\omega-\omega_{n}+i\gamma_{n}/2}.
\end{equation}
Note that the plasmon Green dyadic (\ref{dyadic}) obeys the optical theorem  $\!\int \! dV_{1} \varepsilon''(\omega,\bm{r}_{1}) \bar{\textbf{D}}_{p}^{*}(\omega;\bm{r},\bm{r}_{1})\cdot \bar{\textbf{D}}_{p}(\omega;\bm{r}_{1},\bm{r}')=-4\pi \bar{\textbf{D}}_{p}''(\omega;\bm{r},\bm{r}')$. Correspondingly, the plasmon LDOS,  defined as $\rho(\omega,\bm{r})=-(1/2\pi^{2} \omega) \text{Tr}\, \bar{\textbf{D}}_{p}''(\omega;\bm{r},\bm{r})$,  has the Lorentzian shape 
\begin{equation}
\label{ldos}
\rho(\omega,\bm{r})=\frac{\gamma_{n}}{8\pi^{2} U_{n}} \frac{ \textbf{E}_{n}^{2}(\bm{r})}{(\omega-\omega_{n})^{2}+\gamma_{n}^{2}/4}.
\end{equation}
Frequency integration of Eq.~(\ref{ldos}) yields, with help of Eq. (\ref{energy-LL}), the  plasmon mode density  
\begin{equation}
\label{density}
\rho(\bm{r})=\!\int\! d\omega \rho(\omega,\bm{r})= \frac{2\textbf{E}_{n}^{2}(\bm{r})}{\int \! dV [\partial (\omega_{n}\varepsilon')/\partial \omega_{n}]\textbf{E}_{n}^{2}},
\end{equation}
which describes the spatial distribution of plasmon states and, for typical  $\textbf{E}_{n}(\bm{r})$, represents the inverse plasmon mode volume \cite{lalanne-prl13,maier-oe06}.
Near the resonance  ($|\omega-\omega_{n}|\ll \gamma_{n}$), the plasmon LDOS takes the form
\begin{equation}
\label{ldos-res}
\rho(\omega_{n},\bm{r})=\frac{\textbf{E}_{n}^{2}(\bm{r})}{2\pi^{2} U_{n}\gamma_{n}}=\frac{\textbf{E}_{n}^{2}(\bm{r})}{2\pi^{2} Q_{n}},
\end{equation}
where $Q_{n}$ is given by Eq. (\ref{power-local}) and we used  $\gamma_{n}=Q_{n}/U_{n}$ [see Eq. (\ref{gamma})]. Remarkably, the mode energy $U_{n}$ cancels out, and $\rho(\omega_{n},\bm{r})$ is proportional to the local field intensity normalized by the absorbed power.  In a similar manner, for the CDOS near the plasmon resonance we obtain $\rho(\omega_{n},\bm{r},\bm{r}')=(2\pi^{2} Q_{n})^{-1}\textbf{E}_{n}(\bm{r})\textbf{E}_{n}(\bm{r}')$. Note that we used the real eigenmodes of Eq. (\ref{real}); for local fields in complex form,  Eqs. (\ref{energy-LL}), (\ref{power-local}), (\ref{dyadic}) and (\ref{ldos-res}) (and the above CDOS) are multiplied by $1/2$, but in either case, the  plasmon LDOS  has the universal form (\ref{ldos-main}).

\begin{figure}[tb]
\begin{center}
\includegraphics[width=0.99\columnwidth]{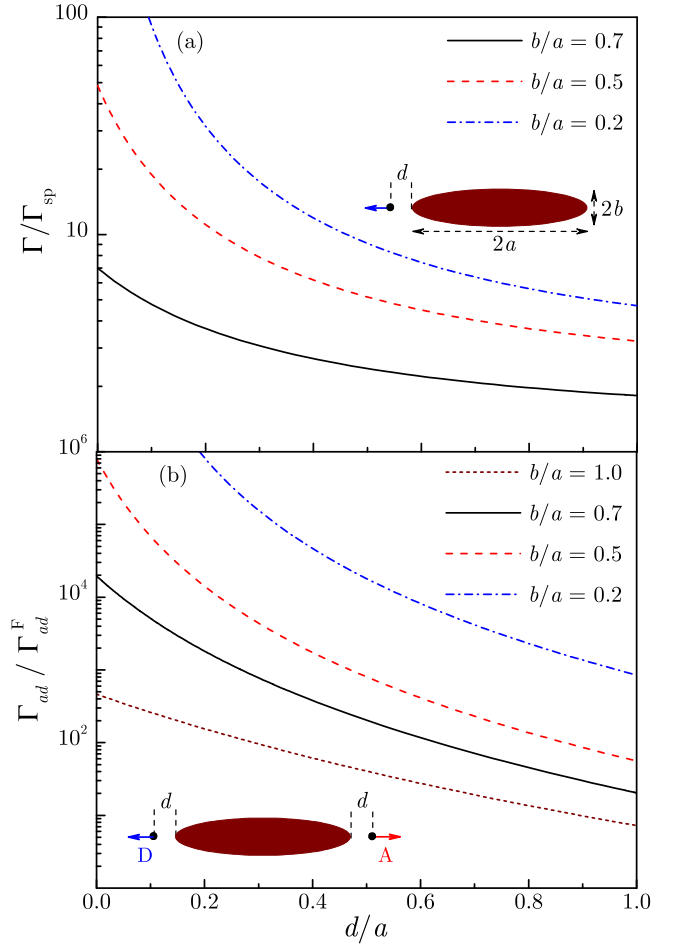}
\caption{\label{fig1} 
(a) Normalized QE-plasmon ET rate and (b) plasmon enhancement  of the F\"{o}rster ET rate for QEs  near the poles of a spheroidal NP with aspect ratio $b/a$. }
\end{center}
\vspace{-6mm}
\end{figure}

\textit{Applications to energy transfer}.---Below, we apply our results to ET between  QEs and plasmons as well as between donors and acceptors  near a plasmonic structure. Consider   a  QE with the dipole moment $\textbf{p}=\mu \bm{n}$ ($\mu$ is the dipole matrix element and  $\bm{n}$ is its orientation) interacting with a resonant plasmon mode [see Fig. \ref{fig1}(a)].  The QE-plasmon ET rate   $\Gamma =(2/\hbar)\text{Im}[\bm{p}^{*} \cdot \textbf{E}(\bm{r})]$, where $\textbf{E}(\bm{r})=- \bar{\textbf{D}}(\omega_{n};\bm{r},\bm{r})\cdot\bm{p}$ is the QE local field,  has the standard form  $\Gamma=(4\pi^{2}\mu^{2}\omega_{n}/3\hbar)\,\bar{\rho}(\omega_{n},\bm{r})$ \cite{novotny-book}, where 
\begin{equation}
\bar{\rho}(\omega_{n},\bm{r})
\!=\!
\frac{-3}{2\pi^{2} \omega_{n}}\,
 \bm{n}\cdot\bar{\textbf{D}}''(\omega_{n};\bm{r},\bm{r})\cdot\bm{n}
\!=\!
\frac{6}{\pi\omega_{n}}
\frac{|\bm{n}\cdot\textbf{E}_{n}(\bm{r})|^{2}}{\int \!dV \varepsilon''|\textbf{E}_{n}|^{2} },
\end{equation}
is the \textit{projected}  plasmon LDOS (hereafter, we adopt   complex field notations), yielding
\begin{equation}
\label{decay}
\Gamma
=\frac{8\pi\mu^{2}}{\hbar}
\frac{|\textbf{n}\cdot\textbf{E}_{n}(\bm{r})|^{2}}{\int \!dV \varepsilon''|\textbf{E}_{n}|^{2} }.
\end{equation}
The rate increases when the losses are reduced, i.e., the plasmon resonance becomes sharper.

To verify Eq.~(\ref{decay}), let us  recover the QE-plasmon ET  rate for a spherical metal nanoparticle (NP) \cite{ruppin-jcp82}. The eigenmodes inside and outside the NP, respectively, have the form $\textbf{E}_{lm}(\bm{r})\propto\bm{\nabla}[r^{l}Y_{lm}(\hat{\bm{r}})]$ and $\textbf{E}_{lm}(\bm{r})\propto a^{2l+1}\bm{\nabla}[r^{-l-1}Y_{lm}(\hat{\bm{r}})]$, where $a$ is the  NP radius, $Y_{lm}(\hat{\bm{r}})$ are the  spherical harmonics ($l$ and $m$ are polar and azimuthal numbers), and the eigenfrequencies $\omega_{l}$ satisfy $l\varepsilon'_{m}(\omega_{l})+l+1=0$. For a QE oriented, e.g.,  normally to the NP surface, we obtain
\begin{equation}
\Gamma_{l}= (2l+1)\frac{(l+1)^{2}}{l\varepsilon''_{m}(\omega_{l})} \frac{2\mu^{2}}{\hbar }\frac{a^{2l+1}}{r^{2l+4}}.
\end{equation}
To illustrate the role of local fields, we plot in Fig.~\ref{fig1}(a) the QE-plasmon ET rate for longitudinal dipole mode in a spheroidal NP normalized by that for spherical NP.

Consider now an \textit{ensemble} of QEs near a plasmonic nanostructure.  The plasmon-induced spatial correlations  between QEs  lead to cooperative effects \cite{pustovit-prl09,pustovit-prb10}, and the ET rates are given by  the eigenvalues of the  decay matrix  $\Gamma_{ij}=(4\pi^{2}\mu^{2}\omega_{n}/3\hbar)\bar{\rho}(\omega_{n};\bm{r}_{i},\bm{r}_{j})$, where  $\bar{\rho}(\omega;\bm{r}_{i},\bm{r}_{j})=-(3/2\pi^{2} \omega) \,\bm{n}_{i}\cdot\bar{\textbf{D}}''(\omega;\bm{r}_{i},\bm{r}_{j})\cdot\bm{n}_{j}$ is the projected CDOS ($\bm{r}_{i}$ and $\bm{n}_{i}$ are, respectively, the QEs' positions and orientations). Using the single-mode chain rule for the CDOS, $\bar{\rho}_{n}(\omega_{n};\bm{r}_{i},\bm{r}_{j})\bar{\rho}_{n}(\omega_{n};\bm{r}_{j},\bm{r}_{k}) =\bar{\rho}_{n}(\omega_{n};\bm{r}_{i},\bm{r}_{k})\bar{\rho}_{n}(\omega_{n},\bm{r}_{j})$, the cooperative  ET  rate  $\Gamma^{c}$ can be   found as 
\begin{equation}
\label{decay-coop}
\Gamma^{c}=\frac{4\pi^{2}\mu^{2}\omega_{n}}{3\hbar}\,\sum_{i}\bar{\rho}(\omega_{n},\bm{r}_{i})
=
\sum_{i}\Gamma_{i},
\end{equation}
where individual rates $\Gamma_{i}$ are given by Eq.~(\ref{decay}). As expected,   $\Gamma^{c}$ scales linearly with the ensemble size.  

We now turn to ET between a \textit{donor} and an \textit{acceptor} located at $\bm{r}_{d}$ and $\bm{r}_{a}$, respectively, near a plasmonic structure  [see Fig.~\ref{fig1}(b)]. The rate of direct (F\"{o}rster) ET due to  donor-acceptor  dipole coupling, $\Gamma_{ad}^{F}$, normalized to the donor radiative decay rate $\gamma_{r}$, has the form \cite{novotny-book}
\begin{equation}
\label{forster}
\frac{\Gamma_{ad}^{F}}{\gamma_{r}}=\frac{9c^{4}}{8\pi }\int \frac{d\omega}{\omega^{4}}f_{d}(\omega)\sigma_{a}(\omega)|T_{ad}^{0}|^{2}=\left (\frac{r_{F}}{r_{ad}}\right )^{6},
\end{equation}
where $f_{d}(\omega)$ and $\sigma_{a}(\omega)$ are, respectively, the donor  spectral function and  the  acceptor  absorption cross section,  $T_{ad}^{0}=-\bm{n}_{a}\cdot\bar{\textbf{D}}_{0}(\bm{r}_{a}-\bm{r}_{d})\cdot\bm{n}_{d}=s_{ad}/r_{ad}^{3}$  is the transition matrix element [$\bm{r}_{ad}=\bm{r}_{a}-\bm{r}_{d}$ is the donor-acceptor distance and $s_{ad}$ is the  orientational  factor], and $r_{F}^{6}=(9c^{4}s_{ad}^{2} /8\pi)\,\! \int\! d\omega f_{d}(\omega)\sigma_{a}(\omega)/\omega^{4}$ defines the F\"{o}rster  distance $r_{F}$ via the QEs' spectral overlap.  The plasmon ET channel is included into Eq.~(\ref{forster}) by replacing $T_{ad}^{0}$  with $T_{ad}=T_{ad}^{0}+T_{ad}^{p}$, where $T_{ad}^{p}=-\bm{n}_{a}\cdot\bar{\textbf{D}}_{p}(\omega;\bm{r}_{a},\bm{r}_{d})\cdot\bm{n}_{d}$ is the plasmon  matrix element  \cite{nitzan-cpl84,dung-pra02,martin-pra03,pustovit-prb11}.  Typically, the QEs' spectral bands overlap well within a  much broader  plasmon band \cite{lakowicz-jf03,andrew-science04,lunz-nl11}, so that  $\bar{\textbf{D}}_{p}$ can be taken at  the resonance $\omega_{n}$. Then, the plasmon matrix element is related to the projected CDOS as $T_{ad}^{p}=(2i/3) \pi^{2}\omega_{n}\bar{\rho}(\omega_{n};\bm{r}_{a},\bm{r}_{d})$, and, using the above chain rule, we obtain the donor-acceptor ET rate as $\Gamma_{ad}=\Gamma_{ad}^{F}+\Gamma_{ad}^{p}$, where
\begin{align}
\label{et-plasmon}
\frac{\Gamma_{ad}^{p}}{\gamma_{r}}
&= \frac{4\pi^{4}r_{F}^{6}}{9s_{ad}^{2}}\,\omega_{n}^{2}\,
 \bar{\rho}(\omega_{n},\bm{r}_{a})\bar{\rho}(\omega_{n},\bm{r}_{d})
\end{align}
is the plasmon-assisted ET rate.  Importantly, $\Gamma_{ad}^{p}$ is proportional to  the  LDOS \textit{product} at the donor  and acceptor  positions and, therefore, exhibits a donor-acceptor symmetry. To gain more insight, let us express $\Gamma_{ad}^{p}$ in terms of individual QE-plasmon ET rates (\ref{decay})   as
\begin{equation}
\label{et-plasmon1}
\frac{\Gamma_{ad}^{p}}{\gamma_{r}}
=\left (\frac{\hbar\Gamma_{a}}{ 2U_{F}}\right )\left ( \frac{\hbar\Gamma_{d}}{2U_{F}}\right ),
\end{equation}
where $U_{F}=\mu^{2}s_{ad}/r_{F}^{3}$ is the dipole interaction at the F\"{o}rster  distance. Factorization of the plasmon-assisted donor-acceptor ET  rate  into the rates of  constituent processes (donor-to-plasmon and plasmon-to-acceptor) reflects the incoherent nature of  ET between different QEs.

%
%


While  F\"{o}rster ET   is   efficient for small donor-acceptor distances,  the system transitions to  a  plasmon-dominated ET regime as $r_{ad}$ increases \cite{lakowicz-jf03,andrew-science04,lunz-nl11}.  The transition onset is reached when $\Gamma_{ad}^{p}\gtrsim\Gamma_{ad}^{F}$, or, using Eqs. (\ref{forster}) and (\ref{et-plasmon1}), 
\begin{equation}
\label{condition1}
\left (\frac{\hbar\Gamma_{a}}{2U_{ad}}\right )\left (\frac{\hbar\Gamma_{d}}{2U_{ad}}\right )\gtrsim 1,
\end{equation}
where $U_{ad}=\mu^{2}s_{ad}/ r_{ad}^{3}$ is the donor-acceptor dipole interaction; i.e.,  when the \textit{widths} associated with   individual ET processes  exceed the direct QE coupling.  The explicit LDOS dependence of the ET rate  allows us to derive, in general form, the plasmon enhancement factor for  F\"{o}rster ET, $\Gamma_{ad}/\Gamma_{ad}^{F}$. After averaging Eq. (\ref{et-plasmon}), i.e., replacing $\bar{\rho}$ with $\rho$ and $s_{ad}^{2}$ with 2/3, and using Eq. (\ref{ldos-main}), we obtain
\begin{align}
\label{condition}
\frac{\Gamma_{ad}}{\Gamma_{ad}^{F}}=1+\frac{3}{2}\left (\frac{ V_{ad}\left |\textbf{E}_{n}(\bm{r}_{a})\right |^{2}}
{\int \!dV \varepsilon''|\textbf{E}_{n}|^{2}}\right )
\left (\frac{ V_{ad}\left |\textbf{E}_{n}(\bm{r}_{d})\right |^{2}}
{\int \!dV \varepsilon''|\textbf{E}_{n}|^{2}}\right ),
\end{align}
where $V_{ad}=4\pi r_{ad}^{3}/3$ is  the spherical volume associated with $r_{ad}$. The ET enhancement factor depends solely on the local field distribution in the system and, therefore, can be varied in a wide range with changing the system shape.

In Fig.~\ref{fig1}(b), we plot $\Gamma_{ad}/\Gamma_{ad}^{F}$ for a donor and an acceptor at a distance $d$ from the opposite poles of a spheroidal NP. As the NP shape changes from a sphere to a thin nanorod, the ET rate increases by several orders of magnitude reflecting the change in the LDOS  that governs the individual QE-plasmon ET rates [see Fig.~\ref{fig1}(a)].

Finally, for ET  between the ensembles of donors and acceptors near a plasmonic structure \cite{pustovit-prb13,poddubny-prb15}, the plasmon contribution to the ET rate factorizes into a product of rates for two constituent cooperative processes: an ET from donors to a resonant  plasmon mode followed by an ET from the plasmon mode to  acceptors. The ET rate between two ensembles is then given by Eq. (\ref{et-plasmon1}), where individual rates $\Gamma_{a}$ and $\Gamma_{d}$ are replaced with their cooperative counterparts  $\Gamma_{a}^{c}$ and $\Gamma_{d}^{c}$, given by Eq. (\ref{decay-coop}).



In summary, the LDOS for any nanoplasmonic system has the universal form (\ref{ldos-main}) in the frequency region dominated by  a plasmon resonance. Explicit formulas, in terms of the plasmon local field, are derived for ET between QEs and plasmons as well as between donors and acceptors situated near a plasmonic nanostructure.  


This work was supported in part by National Science Foundation Grants No. DMR-1610427 and No. HRD-1547754.


\end{document}